\documentclass[aps,pre,preprint,groupedaddress,showpacs,showkeys]{revtex4}
\usepackage{graphics,graphicx}
\usepackage{amsmath}

\begin{document}


\title{Cooperation Evolution in Random Multiplicative Environments.}


\author{Gur Yaari}
\affiliation{The Racah Institute of Physics, The Hebrew University of Jerusalem, Edmond Safra Campus, Givat Ram, Jerusalem, 91905, Israel}
\email[]{gyaari@gmail.com}
\thanks{We acknowledge very instructive discussions with Nadav Shnerb, Ofer Biham, Itzhak Aharon, Damien Challet and Yi-Cheng Zhang. The present research was partially supported by the STREPs CO3 and DAPHNet of EC FP6.}

\author{Sorin Solomon}
\affiliation{The Racah Institute of Physics, The Hebrew University of Jerusalem, Edmond Safra Campus, Givat Ram, Jerusalem, 91905, Israel}
\affiliation{Institute for Scientific Interchange Viale Settimio Severo 65, 10113, Turin, Italy}


\date{\today}

\begin{abstract}
Most real life systems have a random component: the multitude of
endogenous and exogenous factors influencing them result in stochastic
fluctuations of the parameters determining their dynamics.
These empirical systems are in many cases subject to noise of multiplicative nature.
The special properties of multiplicative noise as opposed to
additive noise have been noticed for a long while. Even though
apparently and formally the difference between free additive vs.
multiplicative random walks consists in just a move from normal to
log-normal distributions, in practice the implications are much more
far reaching.
While in an additive context the emergence and survival of cooperation
requires special conditions (especially some level of reward,
punishment, reciprocity), we find that in the multiplicative random
context the emergence of cooperation is much more natural and
effective. We study the various implications 
of this observation and its applications in various contexts.

\end{abstract}


\maketitle
\section{Background and Previous Knowledge}

One of the puzzling facts in game theory is the recurring result,
in a wide range of conditions that the general good is not reached 
by each of the individuals following its own self interest.
Technically it means that in most of the games 
considered, the ``Nash Equilibrium'' is not an optimal situation
for the ensemble of players taking part in the game \cite{Aumann_Maschler1995}.
Another name for this kind of situation is ``the tragedy of the commons''
\cite{Hardin1968}:
if a set of players have access to a common good, and each of them will act 
only in his own interest, the common good is going to be over-exploited and 
ultimately lost$/$destroyed.
The puzzle resides in the fact that the very existence of our society, 
human civilization and many animal societies is based on a much higher 
level of cooperation than suggested by the above models.
The loopholes in the arguments leading to those pessimistic conclusions have
been looked for at length and in certain cases some mechanisms to avoid this 
``mean fate'' were found. For example, allowing for an infinitely long chain 
of iterated games might affect the non-cooperative result \cite{Aumann_Maschler1995}.
Five more mechanisms to ensure that cooperative traits will survive evolution
in additive environments can be found here \cite{Nowak2006}.

In the present paper we offer another alternative: we find that if the 
gains and losses of the participants are multiplicative, then 
the cooperative behavior becomes highly preferable and might in fact 
be realized in a very wide range of realistic examples.
The fundamental feature of multiplicative process we exploit here is
the fact that the expected gain of the players taking part in this 
iterative process depends in a crucial way on the number of players 
considered (number of independent realizations) and the number of 
time steps that the game is played. For long times (the number of time steps 
played in the game), the expected wealth of the players follows the \textit{geometric 
mean} and not the \textit{arithmetic mean} of the game (keep in mind that ``geometric mean $\leq$ arithmetical mean'').

In economics, one way to take into account this effect was to 
declare that what is to be maximized is not the wealth itself but rather 
the ``utility function'' \cite{vonNEUMANN_MORGENSTERN1944}. The case where the
``utility function`` is the logarithm of the wealth $U = \sum_e p(e) \ln(
g(e))$ reduces to considering the geometric mean rather than the arithmetic mean. 
Thus, the use of this utility function may be interpreted as a way to take into account the fact that in general a strategy is applied repeatedly for long spans of time such that the frequency of the events approach their probability. Some of the behavioral anomalies studied over the years \cite{Allais1953,Kahneman_Tversky1979,Thaler1994} can be related to the subtle difference between the expectation for one game and the probability for longer series of events. Moreover, the asymmetry between losing and gaining proposed by the prospect theory \cite{Kahneman_Tversky1979} and the 
particularities of how people evaluate very improbable events get more legitimacy from the above analysis (and the one below).
The present analysis is mathematically congenial with re-balancing strategies \cite {Kim_Markowitz1989} in financial portfolio management. Indeed, one finds there a multiplicative random process (the fluctuations of the equity prices) coupled with a redistribution of the portfolio wealth between the various equities. As in the case of individuals cooperating, the re-balancing of the investments between the various portfolio instruments leads to very crucial optimization \cite{Marsili_ea1998}. The difference is that the portfolio instruments do not have a conscience or personal interests, or the capability to decide by their own what is to happen with the money invested in them. Thus the formal game is quite different: in the present article we ask whether cooperation serves better the selfish interests of the \textit{living} individual while in the portfolio problem one asks what is the strategy for optimizing the total wealth. It is a nontrivial result of the present paper that in a wide range of conditions the two coincide. This is again a nontrivial consequence of the multiplicative dynamics: in usual additive gain$/$loss games one ends up in a globally non-optimize Nash equilibrium. 
The attitude suggested by the analysis that follows is of ignoring events that never happen. It affects crucially the recommended level of sharing in very risky environments \cite{ETHIER2004,Marsili_ea1998}.

In situations that we will study below, this makes a difference of life and death: systems that ''theoretically''
are expected to flourish (arithmetic mean exceeds $1$) may in reality be doomed to extinction (geometric mean is below $1$). 
In order to restore the efficient ``expected`` result, one has to make recourse to extreme forms of sharing that might be difficult to discriminate behaviorally from pure altruism.

\section{The proposed dynamics}
\subsection{The generic setup}\label{generic}

First, in order to emphasize the particularities of the multiplicative random processes, let us evaluate the expected gain $F(T)$ of a multiplicative random process up to time $T$. $F(T)$ is by definition the sum over all possible histories $H$ of duration $T$
of the probability of each history $P(H)$ times the gain $G(H)$ associated with that history: $F(T)= \sum_H P(H)\cdot G(H)$. Each history probability $P(H)$ is the product of the probabilities $p(e_H(t))$ of its events $e_H(t) , t\in\{1,...T\}$:  $P(H) = \prod_t p(e_H(t))$ and the history's gain is, in a multiplicative process, the product of the gains $G(H) = \prod_t g(e_H(t))$ associated with the events. Thus $F(T)$ can be expressed also as:
\begin{equation}\label{eq-1}
F(T)=  [\sum_e p(e)\cdot g(e)]^T 
\end{equation}
However, for large times ($T$) the only histories $H$ appearing with finite probability $P(H)$ are the ones where each event occurs a number of times $ T \cdot p(e)$, proportional to its probability $ p(e)$. Thus, the expected gain instead of being the $T$ power of the arithmetic mean
 $[\sum_e p(e) g(e)]^T$  reduces to $\prod_e g(e)^{p(e) \cdot T}$ i.e. the geometric mean to the power $T$:
\begin{equation}\label{eq0}
 \prod_e g(e)^{T\cdot p(e)} = [\prod_e g(e)^{p(e)}]^T=e^ {T\cdot \sum_e p(e) \ln (g(e))}
\end{equation}

Next, we propose a simple dynamic setup that will allow us to study further the possible implications of this basic feature of multiplicative dynamics.
We consider the following individual iterative process: each player puts all of his/her accumulated wealth $W_i(t)$ at stake :
\begin{equation} \label{proc}
W_i(t+1) = \left\{ \begin{array}{l}
    a \cdot W_i(t) \textrm{  with probability } p  \\
    b \cdot W_i(t) \textrm{  with probability } q=1-p\\
\end{array}\right\}
\end{equation}
where $(0<b<a)$. As was mentioned before, if one takes first the limit of $t \longrightarrow \infty $, then the 
expected wealth of any finite number of \textit{independent} realizations, $R$, will follow the geometric mean ($M_g$), i.e.
for $t \not \ll log(R)$ (see appendix A for the derivation of this condition)
\begin{equation} \label{geometric}
<W_i(t)>_i = <W_i(0)>_i\cdot (M_g)^{t}
\end{equation}
where $M_g\equiv a^{p}b^{q}=e^{p\cdot\log(a)+q\cdot\log(b)}$ and the average is done over the $R$ realizations.
While for any finite time if one takes \textbf{\textit{enough}} \textit{independent} realizations then the expected wealth will follow the arithmetic mean ($M_a$), i.e. when $R\gg e^t$ then
\begin{equation} \label{arithmetic}
<W_i(t)>_i = <W_i(0)>_i\cdot (M_a)^{t}
\end{equation}
where $M_a\equiv p\cdot a+q\cdot b$.
Since it is always true that $M_a \geq M_g$, it turns out that
the na\"{\i}ve (arithmetic) expectation is always deceiving by making
the game seemingly more gainful than it actually is. In the cases where
$M_a \geq 1 \geq M_g$ it becomes a matter of life and death as the arithmetic mean predicts \textit{growth} while
the reality follows the geometric mean which predicts \textit{decay}.

Hereby, we show that there exists a better way
to restore the optimistic expectation value ($M_a$):
it is enough that a small group of individuals share continuously their total wealth to ensure the optimistic expectation value of the game without the need for exponentially large number of realizations.

\begin{figure}[]
\includegraphics[height=3in,angle=270]{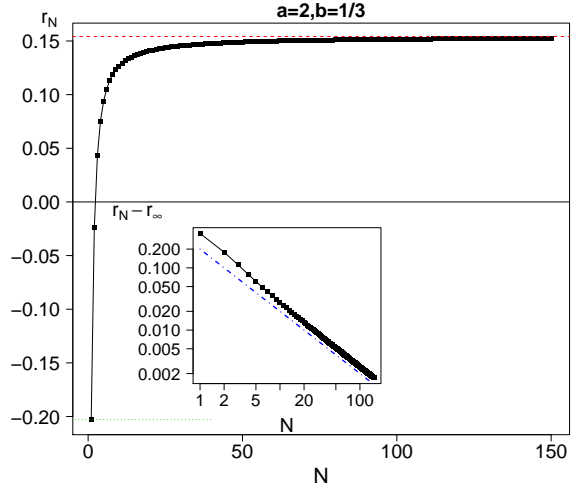}
\caption{ \label{fig1}
The main plot represents the average growth rate $r_N$ as a function of $N$ the number of sharing individuals (according to equation \ref{eq2}). The parameters are ($p=q=0.5,a=2,b=1/3)$ i.e. at each time step one can double ones wealth or loose two thirds of it with equal chances. With these parameters, the geometric mean is less than one ($\sqrt{\frac{2}{3}}$) while the arithmetic mean is larger than one ($\frac{7}{6}$). One sees that the curve $r_N$ crosses $0$ between $N=2$ and $N=3$ which corresponds to $N_{crit}=3$. This means that in order to survive, the individuals have to share with at least two other individuals. The dashed line close to the bottom of the figure is $\ln (2/3)/2$ the value of the logarithm of the geometrical mean of $a$ and $b$ while the dotted line close to the top of the figure is the value of the logarithm of the arithmetic mean $\ln(7/6)$). In the inset, one sees how $r_N$ approaches the value of the logarithm of the arithmetic mean $\ln (7/6)$ as $N$ increases. It shows that $r_N$ approaches the value $7/6$ asymptotically as $1/N$ (dashed line corresponds to $1/N$ dependency).
}
\end{figure}

More specifically, if $N$ players are sharing their gains after each game, the evolution of their wealth will be governed by stochastic process which updates $W_i(t)$ (uniformly to all players) as:
\begin{equation}
W_i(t+1)=[a\cdot \frac{k}{N} + b\cdot \frac{N-k}{N}] \cdot W_i(t)
,\ \ \ \   k\in \left\{0...N\right\}
\end{equation}
with probability $\left( \begin{array}{l}
    N\\
    k\\
\end{array}\right) p^kq^{N-k}$ .
Here $k$ is the number of "wins" ($a$'s) 
the $N$ individuals achieve during the time $t$.
For large enough times ($t\not \ll log(R)$) this leads according to equation \ref{eq0} (Figure~\ref{fig1}) to the expected growth rate of:
\begin{equation}\label{eq2}
r_N=\sum_{k=0}^{N} \left( \begin{array}{l}
    N\\
    k\\
\end{array}\right) p^kq^{N-k}\cdot \ln(a\cdot \frac{k}{N} + b\cdot \frac{N-k}{N})
\end{equation}
where $r_N$ is the exponential growth rate of the new game which is composed of $N$ sharing individuals: i.e. the logarithm of the geometrical mean of the new setup (where the individuals are sharing). Since for large $N \longrightarrow \infty $ the only relevant contribution to this sum comes from $k=p\cdot N$ and the prefactor of the logarithm sums to one (since it is a probability function) one may conclude that:
\begin{equation}
 r_{\infty}= \ln(p\cdot a+ q\cdot b)=\ln(M_a)
\end{equation}

According to the central limit theorem, for large $N$, the variance in the fluctuations of the growth rate around $r_N$ depends on $N$ as $D_N = D/N$ where $D$ is of order $1$ and depends on the parameters $(p,a,b)$. On the other hand, according to the Ito calculus the multiplicative process of the type: 
$V(t+1)=e^\eta\cdot V(t)$ where $\eta$ is a Gaussian variable with an average of $r_N$ and variance of $D_N$ has an
average growth rate equal to $r_N+D_N/2$.  Combining these two observations with the knowledge of $r_\infty$ leads to:

  \begin{equation}
r_N + D/(2N)  \stackrel{_{\longrightarrow} }{_{N \rightarrow \infty}}r_\infty 
 =\log{(p\cdot a+q\cdot b)}
 \end{equation}
and allows one to evaluate the asymptotic dependence of $r_N$ on $N$:
  \begin{equation}
r_N \sim \log{(p\cdot a+q\cdot b)} -D/(2N)  
 \end{equation}
This result is indeed validated by the numerical evaluation of eq. \ref{eq2} as seen in Figure ~\ref{fig1}

\subsection{Specific example}\label{spec}

\begin{figure}[t]
\includegraphics[height=3in,angle=270]{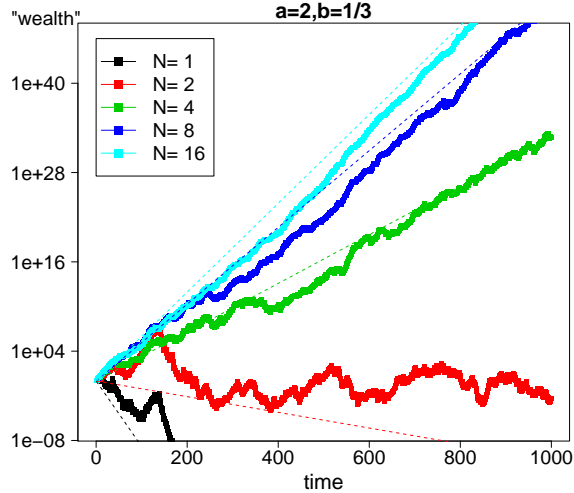}
\caption{\label{fig2}
Comparison between a set of $16$ non-sharing individuals and the same individuals sharing in groups of various sizes ($8$ groups of $2$, $4$ groups of $4$ ... etc). All individuals can double at each time step their wealth or loose $2/3$ of it with equal probabilities (same as in figure ~\ref{fig1}).
Then they share equally the wealth with the partners in their group. One sees that modulo the noise (and some logarithmic corrections), the growth rates approach respectively the theoretical slopes (dashed lines in the figure) corresponding to their respective groupings $r_1, r_2, r_4... $ etc (according equation \ref{eq2}). In order to compare the different results of the groups in a "fair" way, we used the same random numbers for the same individual in the different experiments. The difference was that for the non-sharing individuals we averaged over the $16$ independent realizations of $W(t)$ while for
groups of size $2$  we averaged over the $W(t)$'s of the $8$ sharing pairs, and so on. Note that not only the slopes but also the standard
deviations are different (due to the non-self averaging feature \cite{Derrida1994} of such systems).
}
\end{figure}

In the case where $b \neq 0$ and $M_a>1$ the decay can be avoided completely and forever. In order to build further the intuition for the generic case ($b\neq 0$) let us substitute the numbers $p=q=0.5$, $a=2$ and $b=1/3$ in the general process defined by equation \ref{proc}. Here we have: $M_a\equiv7/6>1>M_g\equiv(2/3)^{(1/2)}$ For these parameters according to equation \ref{eq2}, $r_2 < 0< r_3 $ and thus $N_{crit} =3$ .
When a group consist of $  N = N_{crit}$ individuals, every and each one of these individuals cannot afford to defect since then the rest of the population will decay and even if the others would keep sharing with him/her, the entire system will eventually collapse, including the defector. One could argue that this argument holds only in cases where $N = N_{crit}$ and when $N > N_{crit}$ one can still decide to defect and do better than the rest. Our answer to this argument may sound similar to group selection arguments but with a slight change: we have showed that the asymptotic growth rate is an increasing function of $N$. Hence, on average, each of the individuals in a group with $N$ individuals with one defector will have the effective growth rate
of individuals in a group with $N-1$ individuals. In principal, if this rate is still positive, one may be tempted to defect, but in practice, if this group is part of a larger population, each of the individuals will do worse (on average) than other individuals from other groups with $N$ \textit{sharing} individuals. Thus if one introduces competition for limited resources, having a lower growth rate means having a lower chance to survive, even if the competition is still on an \textit{individual level}. Thus the sharing behavior is stable against defection in so far as it is in the interest of each of the individuals not to defect.
 See Figure~\ref{fig1} for plots of the numeric solutions for these parameters and Figure~\ref{fig2} for the validation of these solutions with stochastic simulations.

\section{The Kelly terminology}

\begin{figure}[t]
\includegraphics[width=3in,angle=0]{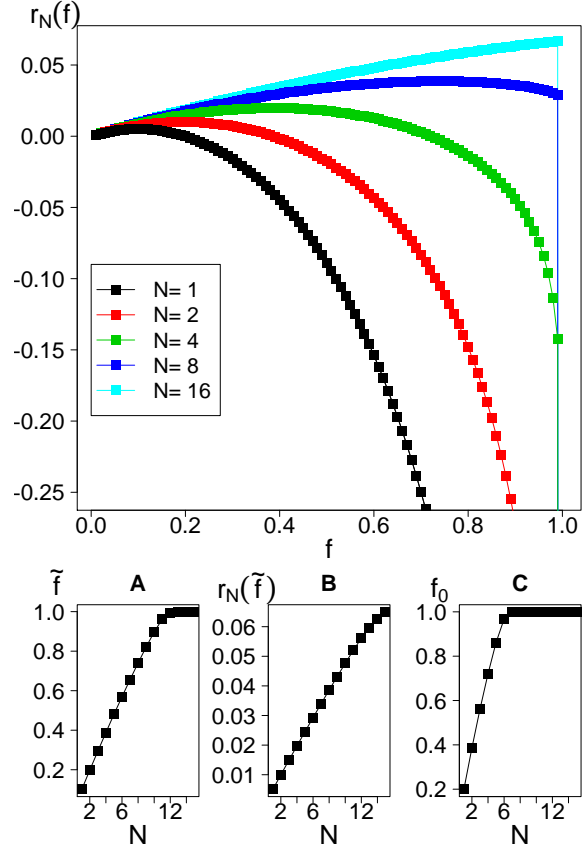}
\caption{\label{fig3}
The curves in the main figure represent $r_N$ ($N=1,2,...$) (according equation \ref{equ9})
as a function of $f_{Kelly}$ (the gambled fraction, $f$ of one's current wealth) 
for a game with $d=1$ (i.e. getting $2\$$ on each $1\$$) and $p=0.55$. One sees how the range of $f$ where $r_N$ is negative
shrinks extremely fast as a function of $N$. Thus, for large enough $N$ sharing individuals, almost any $f$ is safe even if one does not know the parameters of the game ($d$ and $p$). Inset $C$ shows the boundary of the safe $f$ region as $N$ increases, Inset $A$ shows the position of the optimal $f$ as a function of $N$. Inset $B$ shows that the maximum of $r_N$ (equation \ref{equ9}) increases as $N$ grows: the optimal Kelly strategy for $N$ sharing players improves with $N$.
}
\end{figure}

As seen above, high risk, high gain potential processes generate a paradox: the naive expected wealth grows asymptotically to infinity while the probability 
for non-vanishing wealth (the survival probability) decreases exponentially. To overcome this puzzle, J. Kelly,\cite{Kelly1956} initiated an approach that led to a vast literature \cite{ROTANDO_THORP1992,THORP1969,VINCE1992,MILLER1975,MARKOWITZ1976,ETHIER2004,BROWNE2000,AUCAMP1993,ALGOET_COVER1988,Medo_ea2008} that tested, extended 
and applied practically his proposal. The mathematical setting that Kelly considered was the following. Each individual plays an iterated chance game where at each time step $t$ the player might loose his/her (whole) wealth with probability $q$ or gain a fraction $d$ of it with probability $p=1-q$. The player can decide which fraction $f$ of his current wealth $W_i(t)$ to stake. According to these rules, at the end of the time step $t$, the player's wealth will become 
\begin{equation}\label{Kdyn}
W_i(t+1) = \left\{ \begin{array}{l}
    (1+fd)  W_i(t) \textrm{with prob.} p \\
    (1-f)  W_i(t) \textrm{with prob.} q=1-p\\
\end{array}\right\}
\end{equation}
Kelly claimed that for asymptotic times and fixed $f$ the individual wealth behaves like
$W_i(t)\sim e^{t[p\cdot \ln(1 +f\cdot d) + q\cdot \ln(1-f)]}$ (which is equivalent to our notation of $(M_g)^t$). Thus, he concluded that the quantity to be maximized is:
\begin{equation}\label{kelly}
r = {[p \cdot \ln (1 +f d) + q \cdot \ln (1-f) ]}
\end{equation}
 which leads to an optimal value for $f$, $f_1\equiv \frac{(pd-q)}{d}$.
Note that for each choice of $f$ the Kelly setup can be connected to the multiplicative random walk formulation by substituting in \ref{proc}: 
$a=(1+fd)$ and $b=(1-f)$. In these terms, the Kelly criterion is an expression of the fact that for large enough times,
the actual behavior of one individual wealth is determined by $r_1$ as explained in equation \ref{geometric}.
Thus the Kelly criterion is indeed relevant to our problem: it finds the $f$ value that maximizes $r_1$.
One may combine the Kelly prudence $f\neq1$ strategy with the sharing strategy leading to equation \ref{eq2} in order to further optimize performance as we show below. Thus in the light of the previous section results one can address a fundamental subject that the Kelly criterion raises: In real life, one often does not know for sure what game he/she is playing or if the game is not changing in time. Therefore one is not sure which Kelly strategy ($f$) to choose: in case the $f$ that had been chosen gives negative $r_1$ for the actual game played ("the reality") 
one is in danger to go bankrupt. The way to cope with this high level of uncertainty is \textit{to share}: 
If a group of $N$ individuals decide to share their wealth and redistribute it at each time step, then they will have more chances to have 
positive $r_N$ for a wider range of different games ($(p,d,f)$). More precisely, not only that the long run expected gain increases 
with the size of the group but also the domain where $r_N$ is negative: 
\begin{equation}\label{equ9}
 r_N=\sum_{k=0}^{N} \left( \begin{array}{l}
    N\\
    k\\
\end{array}\right)p^kq^{N-k} \ln((1+df) \frac{k}{N}+(1-f) \frac{N-k}{N})<0
\end{equation}
shrinks significantly. For $N$ of order $10$ (and $p=0.55,d=1$), the region of ''dangerous`` $f$'s shrinks from $0.8$ (out of $1$) to $10^{-20}$ !! (Figure~\ref{fig3},inset \textbf{C}). This fact allows the player not only to win more asymptotically, but also to shrink significantly the risk one takes!
These results are presented in Figure ~\ref{fig3}. 

\section{Effects of asynchronous updating:
infinite survival time for games with total loss}
\begin{figure}[t]
\includegraphics[width=3in,angle=0]{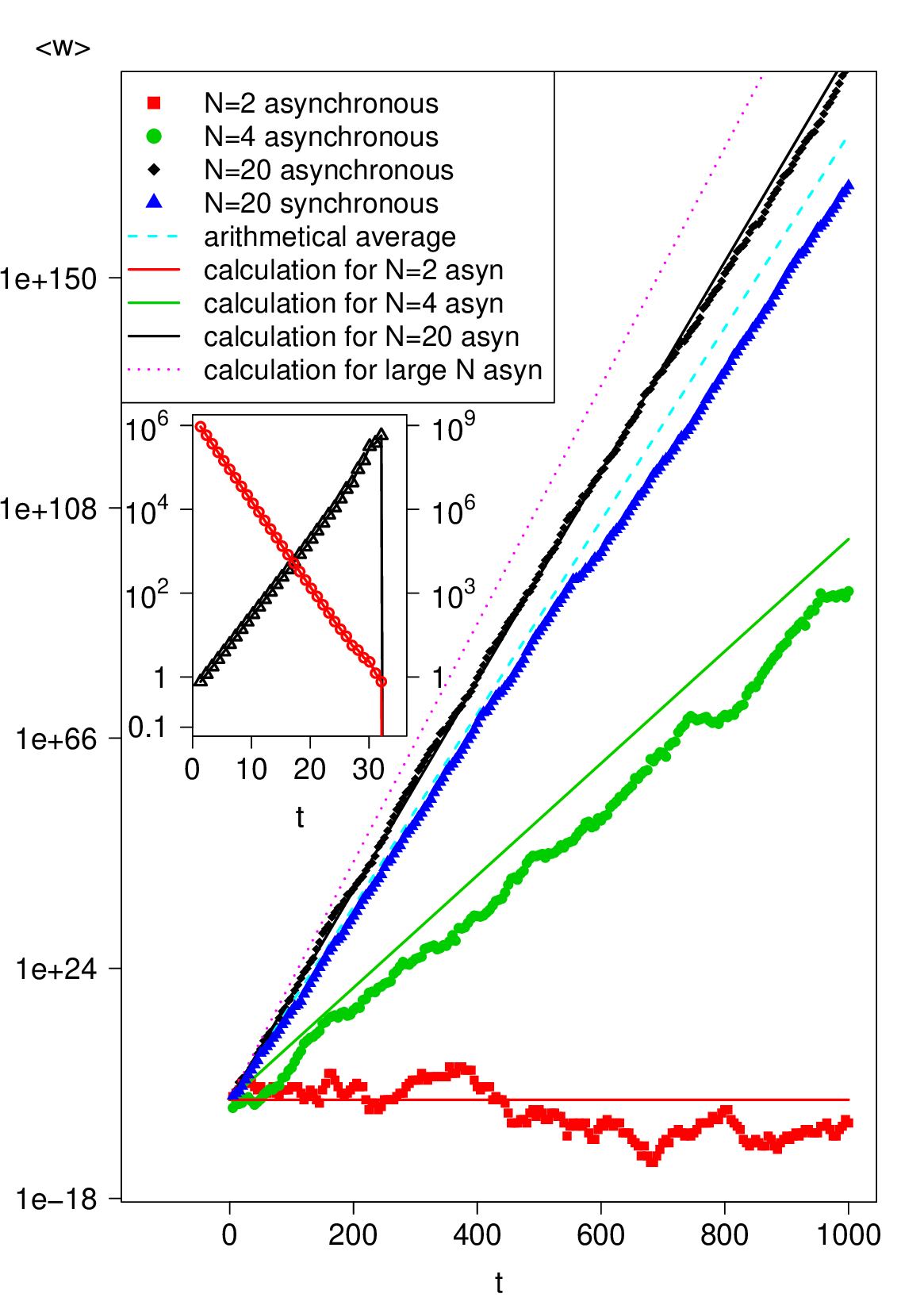}
\caption{\label{fig4}
The evolution of $W(t)$ for synchronous and asynchronous updated sets of sharing individuals.
The scale is semi-logarithmic meaning that straight lines correspond
to exponential.
All series fit well the straight line with the theoretically predicted
slope. In principle the synchronous run should collapse suddenly
and irreversibly to $W(t) =0$ when by chance all individuals
have bad luck (loose all their current individual wealth) \textit{at the same time step.}
However this very improbable event is expected
only around times of order $2^{20}$ which are beyond the size of
the runs shown in the figure ($T=1000$).
Compare this with the case of non-sharing
individuals in the inset where even for $N=10^9$ the collapse takes place after
only $\ln (10^9)\sim 20$ steps! The (red) circles in the inset represent the number of $N(t$) individuals with $W_i (t) \> 0$
(in fact all the $W_i (t)$ which are not zero are $3^t$, the corresponded scale is the right one).
The (black) triangles represent the average $W(t) = N(t)/N(0) 3^t$.
Note that $N(t)$ decays as $2^{-t}$ and thus $W(t)\sim (3/2)^t$ until
$N(t) =0$ when $w(t)$ collapses to $0$ too! The life time of this process
is exceedingly short: in the presented realization, even for $N(0)=10^9$, $W(33)= 0$.
Returning to the main plot, the diamonds correspond to a $N=20$ asynchronous sharing individuals.
The slope fits well the straight line with the theoretical value of the
slope $r=10 \cdot \ln(1.045)$, which is clearly above the maximal
value allowed for synchronous dynamics $r= \ln(3/2)$ but below the line with asymptotic
$N=\infty$ slope $r= 1/2$.
}
\end{figure}

When the system has some kind of external clock one may treat it
in a synchronous fashion: i.e. all agents update their endowments 
in the same instance. In many other systems the individuals are not 
updated simultaneously but rather asynchronously. 
The theoretical framework presented so far applies well to asynchronous systems as well.
For instance consider $N$ individuals with wealths $w_i ; i=1,...,N$
who are forced to play each in his turn the following multiplicative 
random game:
We assume that the time advances in units of $\frac{1}{N}$.
Every time unit we choose an agent $i$ randomly and update its wealth according to:
\begin{equation} \label{eq8}
W_i (t+ \frac{1}{N}) = \left\{ \begin{array}{l}
    a  \cdot W_i(t) \textrm{  with prob. } p  \\
    b  \cdot W_i(t) \textrm{  with prob. } q \textrm{  }(b<1)\\
\end{array}\right\}
\end{equation}
Before and after each such step, the total wealth is redistributed equally between all $N$ individuals.
Thus $W(t)$ follows dynamics where only a fraction $\frac{1}{N}$ is at stake in the 
game while a fraction $\frac{N-1}{N}$ is passive:
\begin{equation} \label{eq9}
W (t+ \frac{1}{N}) = \left\{ \begin{array}{l}
    (\frac{1}{N}  a  + \frac{N-1}{N} ) \cdot W(t) \textrm{  with prob. } p  \\
    (\frac{1}{N}  b  + \frac{N-1}{N} ) \cdot W(t) \textrm{  with prob. } q\\
\end{array}\right\}
\end{equation}
This representation reduces to the Kelly dynamics, of one player described by equation \ref{Kdyn} with
$f = \frac{1-b}{ N}$ and $d= \frac{a-1}{1-b}$.
Thus the growth (/decay) rate of this process is cf. \ref{kelly}
\begin{equation}\label{equ12}
\begin{array}{l}
r^{async}_N = < N \ln (\frac{W(t+1/N)}{W(t)})>= \\
    ={N [p \cdot \ln (1 +\frac{a-1}{N}) + q \cdot \ln (1- \frac{1-b}{ N}) ]}
\end{array}
\end{equation}
where we took into account that the time interval between two individuals 
updating is $\frac{1}{N}$ rather then $1$ which was the time for updating all the $N$
individuals.
Note that for $N \longrightarrow \infty$ one gets:
$r^{async}_{\infty}= p\cdot a +q\cdot b -1$ 
which is larger then the ideal synchronous result implied by equation \ref{arithmetic}:
$r_{ideal} =  \ln (M_a)$
that seemed to be the maximum that one can aspire to. 
The various values for asynchronous processes with $a=3, b=0 , p=q= 1/2,
N= 2 , 4 , 20$ are shown in figure~\ref{fig4}.
One sees that already for $N=2$ one avoids decay
 ($r_2 = 0$).
See inset in figure~\ref{fig4} for the non-sharing result with $N= 10^9$: 
in the presented realization, all the billion non-sharing individuals die after $32$ steps !!
For $N=20$ sharing, synchronous individuals one gets close to the 
ideal average $r_{ideal}=\ln (\frac{3}{2})$ (modulo disastrous collapses 
after about $2^{20} \sim 10^6$ steps which is not seen for 
runs of order $T=1000$ as in figure~\ref{fig4}).
Asynchronous $N=20$ sharing individuals 
exceed the na\"{\i}ve average rate ($\ln (\frac{3}{2})$) and equals
$r^{async}_{20} = {10 [\ln (1.1) + \ln (0.95) ]}$  (figure~\ref{fig4}).

\section{Fixed group size and varied generosity.}

Let us now fix the group size $N$ and look at the asymptotic 
growth rate. In particular, we look at the case in which after
each timestep one is donating to the other members of the group
a fraction $0 \geq D \geq 1$ of the difference between one's wealth 
and the average wealth of the group. In this way the accumulated wealth 
of the group is unharmed after this sharing step.

\begin{figure}[ht!]
\includegraphics[height=3.in,angle=270]{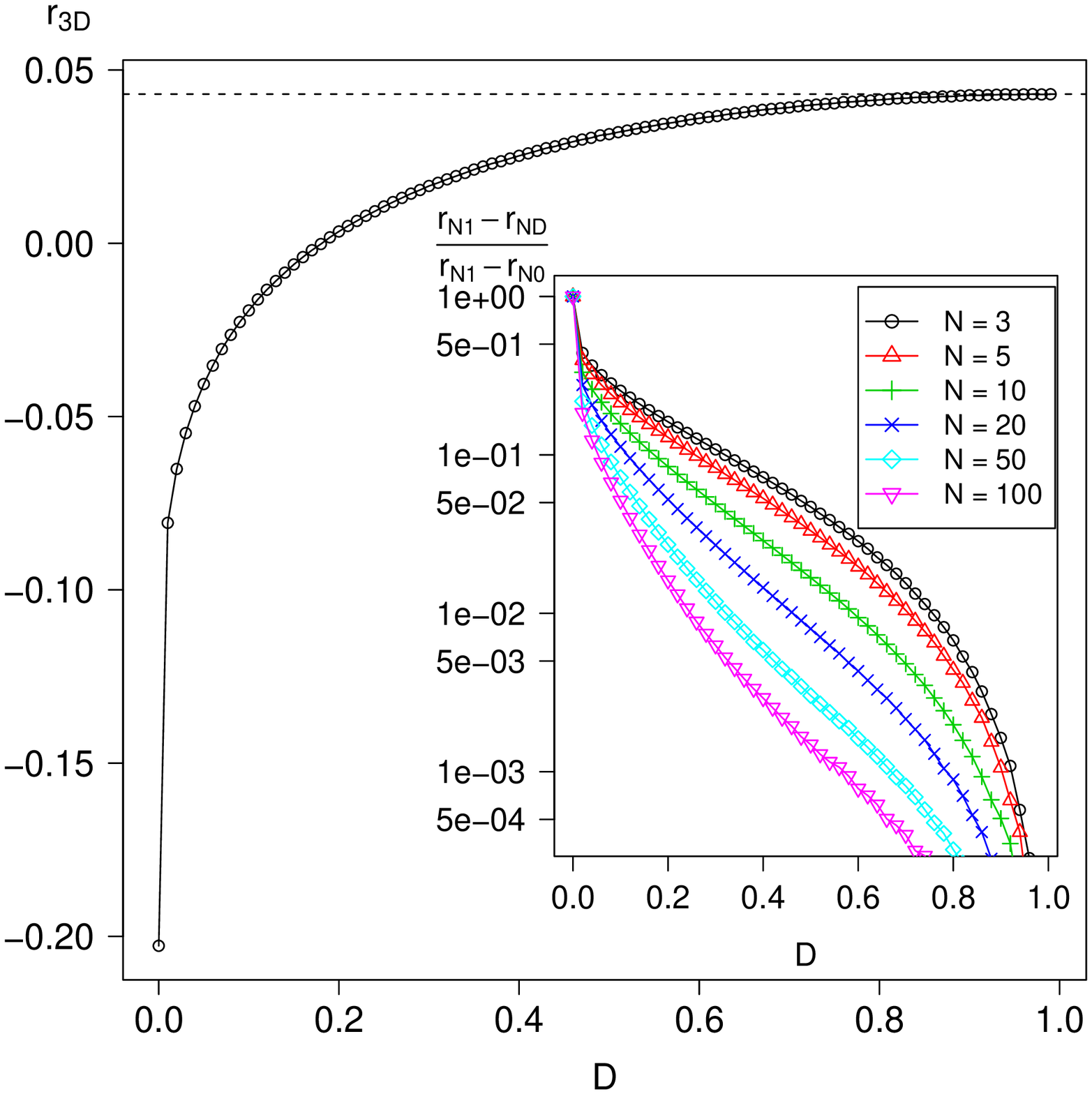}
\caption{\label{fig5}
Same parameters as in figures ~\ref{fig1} and ~\ref{fig2} (and also in figure ~\ref{fig6}, $p=q=0.5$, $a=2$ and $b=1/3$). In this plot we show how the asymptotic growth rate of a group of $N=3$ cooperative individuals approaches the calculated value for $D=1$ (strait dashed line). In the inset the value of $D=1$ minus the asymptotic growth rate (normalized to the value of $D=0$) is plotted on a logarithmic scale for different group sizes. One sees how it approaches the $D=1$ value: the growth rate approaches the $D=1$ value faster as the group size increases.
}
\end{figure}

What we see is that the asymptotic growth rate approaches the calculated value for $D=1$ (absolute redistribution) pretty quickly (inset in fig. ~\ref{fig5}). As one can see in the inset, the rate in which the asymptotic growth rate approaches the $D=1$ value changes with the group size $N$, the bigger it is the less one needs to donate in order to get closer to the $D=1$ value (this is on top of the fact that the $D=1$ value grows as $N$ grows).

\begin{figure}[ht!]
\includegraphics[height=3in,angle=270]{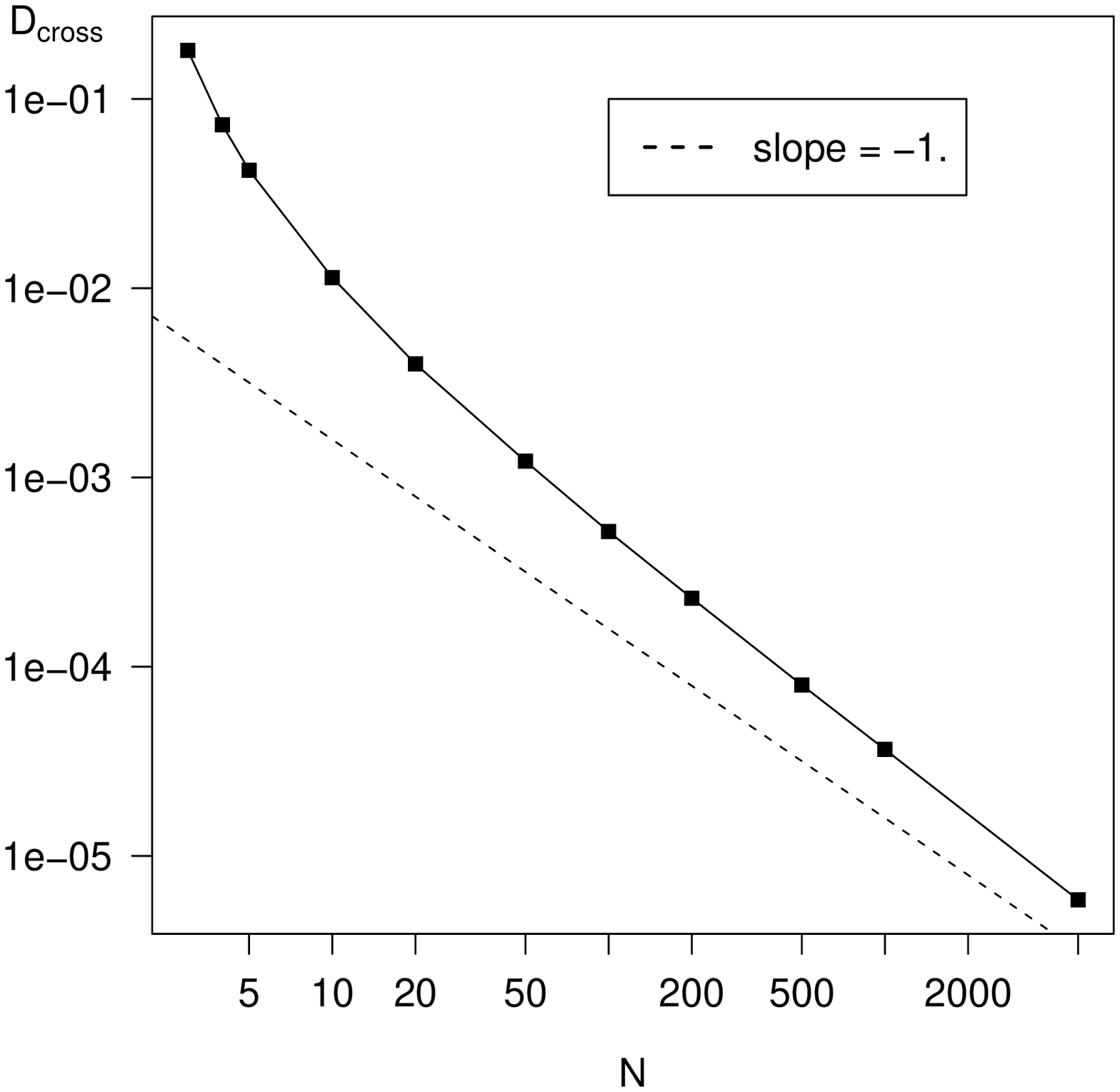}
\caption{\label{fig6}
Same parameters as in figures ~\ref{fig1} and ~\ref{fig2} and ~\ref{fig5}. The value of $D$ that makes the the growth rate of $N$ cooperative individuals positive ($D_{crit}$), on a log-log scale. One sees that asymptotically, it seems to scale like $1/N$ }
\end{figure}

If one is interested in a certain value that the asymptotic growth rate passes (for example $0$, which distinguishes between growth and
decay), it turns out that as the critical value for $D$, $D_{cross}$ scales like $1/N$ for large $N$'s (figure ~\ref{fig6}) which means that as group size increases one has to share less in order to achieve growth rates which are close to the $D=1$ value. 
%
%

\section{Discussion}

Altruism elicits in humans very powerful and diverse feelings. Its paradoxical nature makes it mysterious and challenging to understand. In fact, often one looks for hidden ulterior motives that would offer alternative explanations for an ostensibly altruistic behavior. Indeed, it is difficult to understand why a rational being would sacrifice its own interests or existence for somebody else. At the biological level, it is difficult to understand how a gene that is not selfish would be able to survive and procreate. In a world of scarce resources, forfeiting one's own means for the benefit of another seems synonym with suicide. Thus one often recurs for an explanation to assuming either kinship or some kind of reciprocity. Reciprocity that cannot be enforced seems to be ruled out: whether one is granted a ''gift'' from the other players or not, the formal game-theory optimal reaction is selfishness. Most of the mechanisms that could lead to altruism are marred by the ''free riders'' problem. We trace the origin for the current direction the scientific discussion about altruism is taking to the fact that the vast majority of models considered in the literature\cite{Nowak2006,Lehmann_Keller2006} have an additive reward mechanism. While additive rewards exist in nature in some cases, multiplicative processes are as important and present in many biological and economic systems in various contexts. We have showed that in quite realistic conditions (multiplicative reward mechanism) the very selfish
interest of individuals to survive and prosper leads them rationally to an altruist-like behavior. More precisely, it follows from the rules of the game that by granting unconditionally to others part of one's own resources, individuals prolong their own survival expectation and increase their ``wealth'' expectation. In an additive environment the maximum that could be obtained was the prevalence of tit-for-tat-like \cite{Axelrod_Hamilton1981,Lehmann_Keller2006} behavior that is significantly less then unconditional cooperation. For infinite or indefinite game durations the emergence of cooperation has been obtained within the usual probabilistic $/$ game theory formalism. However, we find that in multiplicative environments (which we believe are the more common scenarios present in the real world), the cooperation is justified and in fact unavoidable even for series with a quite limited number of iterated games. As seen in figure~\ref{fig1}, the difference between cooperation and defection may be the difference between immediate extinction to unlimited growth. More research about the exact conditions in which altruism is preferred (and to what level) in multiplicative random environments is still needed and the results may depend on the specific details of the system such as: what is the exact sharing mechanism? what fraction of all resources that were intended to be shared reach their target and what gets lost along the way? with whom does the group of sharing individuals compete? what exactly is the selection mechanism, etc. However, we may now start to understand why the present surviving species appear to contain altruistic genes:
The genes are still selfish but the phenotype behaves ostensibly altruistic.

\section{Conclusions}

We have studied some aspects of the emergence, efficiency and consequences of cooperation in systems with multiplicative gains and losses. 
We have found that as opposed to games with additive gains and losses, in a risky multiplicative environment unconditional cooperation is quite 
a normal outcome. This conclusion is robustly resilient even when one relaxes and extends the conditions and the rules of the game in various directions.
While certain mathematical properties of multiplicative random walks \cite{Kelly1956,Redner1990,Huang_Solomon2001a,Blank_Solomon2000} underlying our analysis have been known for a while (especially in the context of wealth redistribution between portfolio items for portfolio optimization by the portfolio owner) the application to independent rational individuals each of which decides independently on the basis of self-interest has not been discussed before. The present article is only a modest beginning in a quite wide range of possible research works to explore the rich and unexpected properties of such a multiplicative games theory. In particular, the multiplicative property implies the coupling of the gains in the future to the ones in the past which is an element absent in the additive game theory and had to be substituted through the introduction of memory and "moral" related behavior and additional concepts: reward, punishment, etc. Obviously, when the empirical system follows an additive dynamics, one is to study it through additive rewards games. There are plenty of examples for real life systems that do not fall into this category and where the rewards are of multiplicative nature. We hope that in these cases people will start to study these systems with the appropriate underlying dynamics and will apply the results presented here (and the ones to follow) to them. 

\appendix
\section*{Appendix A: Average behavior vs. typical one}\label{typical}
Let us show why the restoration of the arithmetic means requires exponential number of realizations:
to start with, we concentrate on the case where: $a \geq 1 \geq b \geq 0$ and $M_a>1>M_g>0$
The condition that a realization has a positive
cumulative growth factor 
\setcounter{equation}{1}
\begin{equation*}\label{condition}
\tag{A.1}
\frac{W(t)}{ W(0)} = a^n \cdot b^{t-n} > 1
\end{equation*}
is that the number of ``wins'' ($n$) will be
\begin{equation*}
\tag{A.2}
n > \alpha t 
\end{equation*}
where  $\alpha = 1 /(1-\frac{\ln( a)}{ \ln (b)}$).\\
Given the Gaussian approximation of the Poisson probability of $n$ "wins" (Figure ~\ref{fig7}):
\begin{equation*}
\tag{A.3}
Prob(n,t) = {(2\pi t pq)}^{-1/2} e^{-(n-tp)^2/2tpq}
\end{equation*}
one can compute the fraction of realizations of duration $t$ 
for which the cumulative wealth \textit{grows}
$W(t) / W(0)  > 1$:
\begin{equation*}
\tag{A.4} 
\int_{n=\alpha t}^t {(2\pi t pq)}^{-1/2} e^{ -(n-tp)^2 / 2tpq} dn
\end{equation*}
It turns out that for $M_g <1$, the integral is very well approximated 
by its largest integrand $e^{-(\alpha t-tp)^2 / 2tpq}$.
Thus the measure of the increasing $W(t)$ histories decays to $0$ as:
\begin{equation*}
\tag{A.5}
Prob (W(t) /W(0) >1) \sim e^{-t\{1/(1-\frac{\ln (a)}{\ln (b)})-p\}^2 /2pq} 
\end{equation*}
Which means that in order to have any chance to obtain a result for which $W(t)$ increases at all  
one needs a number of realizations diverging exponentially with $f(a,b,p)\cdot t$. For $t$ time steps:
\begin{equation*}
\tag{A.6}
R \sim e^{ t \{1 /(1-\frac{\ln(a)}{\ln(b)}) -p\}^2 /2pq}
\end{equation*}
For any finite sample $R$, the actual behavior in most
cases will be decay.
One can replace equation \ref{condition} with a more general condition which requires contributions larger than the arithmetic mean:
\begin{equation*}\label{condition2}
\tag{A.7}
\frac{W(t)}{ W(0)} = a^n \cdot b^{t-n} > M_a^t
\end{equation*}
which yields 
\begin{equation*}
\tag{A.8}
R \sim e^{ t\{\frac{\ln(M_a/b)} {\ln (a/b)}-p\}^2 /2pq}
\end{equation*}
that generally means that the number of realizations needed to restore the (arithmetic) average of a game grows exponentially in $t$ - the rate of convergence of the exponential dependency in $t$ is a function of the parameters of the game $(a,b,p)$:

\begin{figure}[t]
\includegraphics[width=3in,angle=0]{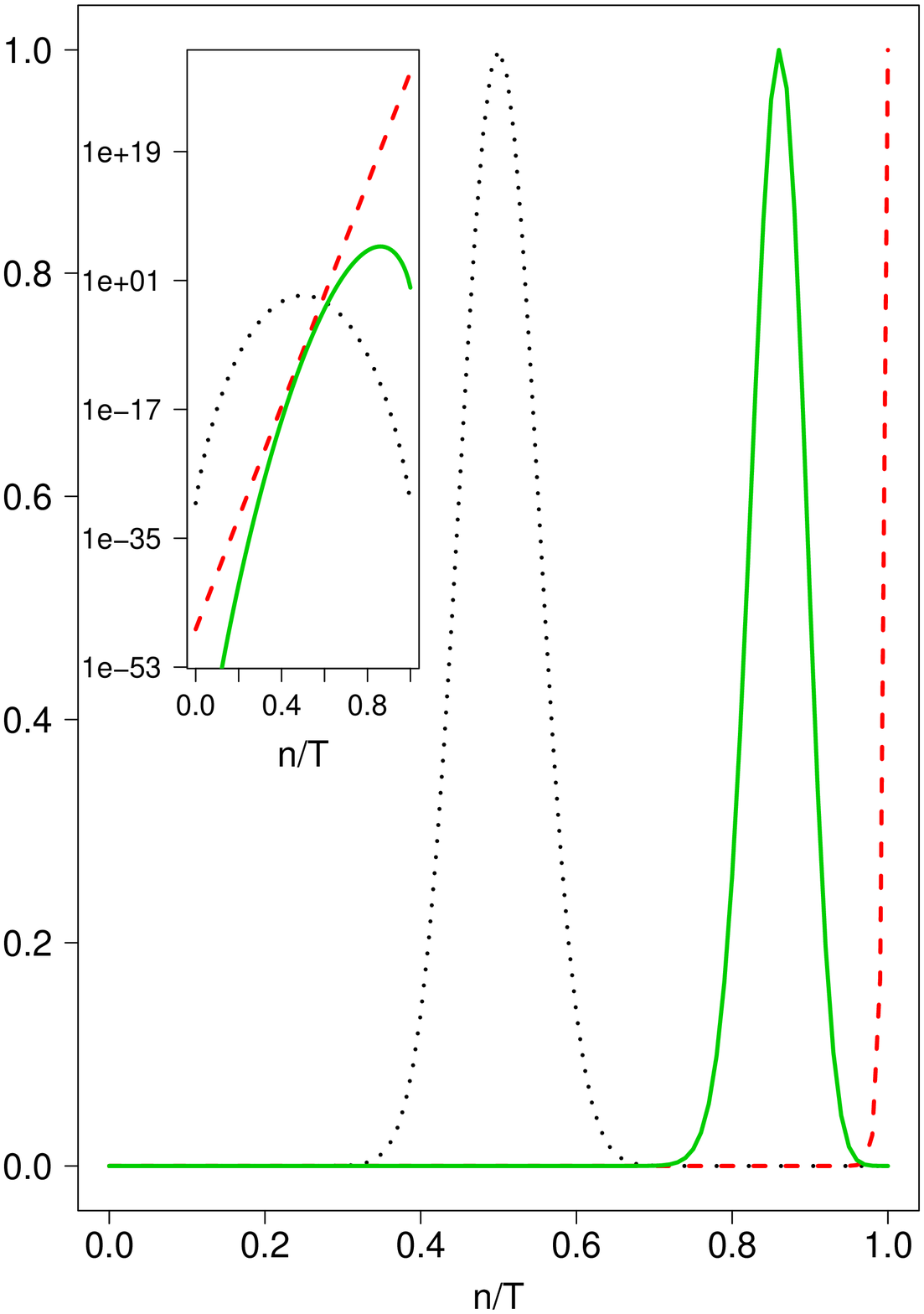}
\caption{\label{fig7}
The contribution of the various types of processes to $W(T)$
as a function of the fraction of ``wins'' $n/T$.
We took $T=100, a=2, b=1/3 , p=q=1/2$ as an example.
The curves represent respectively:
 $-$ the dashed line is the value of $W(T=100)$ for $n$ wins and $100-n$ losses: $2^n 3^{n-100}$.
 $-$ the dotted line is the probability for $n$ wins and $100-n$ losses: $C(100,n)/2^{100}$.
 $-$ the full line is the total contribution to $<W(T)>$ of processes with $n$ wins and $100-n$ losses: $C(100,n) /2^{100} \cdot 2^n 3^{n-100}$.
On the logarithmic scale (inset) the situation seems simple:
the first curve is a straight line, the second is a parabola and the third is
their sum: a parabola with shifted maximum.
However, on the normal graph (the axes are rescaled to facilitate view)
the root of the various paradoxes becomes apparent: the first $2$ graphs
(dashed and dotted) are concentrated in narrow respectively disjoint regions: 
the region where the large values of $W$ are concentrated have negligible probability.
Thus their product (full line) has a spurious
maximum in a region that is practically unreachable in the physical reality.
The relevant values in reality are the ones which take into account only the
region around the peak of the dotted line at $n=50$ where the contribution to the value of $W(T)$ is $(\frac{2}{3}) ^{50}$.}
\end{figure}

\end{document}